\documentstyle[prd,preprint,tighten,aps,amssymb,newlfont]{revtex}
\begin{document}
\draft
\preprint{
\begin{tabular}{r}
\null \vspace{1cm} \null\\
DFTT 05/98\\
hep-ph/9802201\\
\null \vspace{1cm} \null
\end{tabular}
}
\title{Implications of CHOOZ results
for the decoupling of solar and atmospheric neutrino oscillations}
\author{S.M. Bilenky}
\address{Joint Institute for Nuclear Research, Dubna, Russia, and\\
INFN, Sezione di Torino, and Dipartimento di Fisica Teorica,
Universit\`a di Torino,\\
Via P. Giuria 1, I--10125 Torino, Italy}
\author{C. Giunti}
\address{INFN, Sezione di Torino, and Dipartimento di Fisica Teorica,
Universit\`a di Torino,\\
Via P. Giuria 1, I--10125 Torino, Italy}
\maketitle
\begin{abstract}
We have considered
the results of solar and atmospheric neutrino oscillation experiments
in the scheme of mixing of three neutrinos
with a mass hierarchy.
It is shown that
the recent results of the CHOOZ experiment
imply
that
$|U_{e3}|^2\ll1$
($U$ is the neutrino mixing matrix),
that the oscillations of solar neutrinos are described
by the two-generation formalism
and that
the oscillations of solar and atmospheric neutrinos decouple.
It is also shown that
if not only
$|U_{e3}|^2\ll1$
but also
$|U_{e3}|\ll1$,
then
the oscillations of atmospheric neutrinos do not depend
on matter effects and are described by the two-generation formalism.
In this case,
with an appropriate identification of the mixing parameters,
the two-generation analyses of solar and atmospheric neutrino data
provide direct information on the mixing parameters
of three neutrinos.
We discuss the possibility to get information
on $|U_{e3}|^2$
in long-baseline
neutrino oscillation experiments.
\end{abstract}


\section{Introduction}
\label{Introduction}

The existence of the solar \cite{Berezinsky}
and atmospheric \cite{Gaisser}
neutrino problems
have received in the last months impressive
confirmations from the preliminary
results of the SuperKamiokande experiment
\cite{SK-sun,SK-atm}.
The traditional analyses of
solar
\cite{Homestake,Kam-sun,GALLEX,SAGE,SK-sun}
and atmospheric
\cite{Kam-atm,IMB,Soudan,SK-atm}
neutrino data in terms of neutrino oscillations
(see \cite{BP78,BP87,CWKim})
have been performed under the assumption of two-neutrino mixing.
In particular,
this assumption has been adopted in
the most recent analyses of the solar \cite{HL97,FLM97}
and atmospheric \cite{Valencia}
neutrino data,
which include preliminary data from SuperKamiokande
\cite{SK-sun,SK-atm}.
However,
we know that there are three light flavor neutrinos,
$\nu_e$, $\nu_\mu$, $\nu_\tau$, 
that can participate to the oscillations of
solar and atmospheric neutrinos
and it is reasonable to ask what information on the
mixing of three neutrinos
can be obtained from
the two-generation analyses of solar and atmospheric data.

Here we consider a scheme of mixing of three massive neutrinos
with the mass hierarchy
\begin{equation}
m_1 \ll m_2 \ll m_3
\,,
\label{01}
\end{equation}
which is motivated by the see-saw mechanism \cite{see-saw}
and by the analogy with the mass spectra of
charged leptons and up and down quarks.
The masses $m_k$ ($k=1,2,3$) are
associated to the massive neutrino fields
$\nu_k$,
whose left-handed components $\nu_{kL}$
are connected with the
left-handed components $\nu_{{\alpha}L}$ ($\alpha=e,\mu,\tau$)
of the flavor neutrino fields
by the relation
\begin{equation}
\nu_{{\alpha}L}
=
\sum_{k=1}^{3}
U_{{\alpha}k} \, \nu_{kL}
\,,
\label{00}
\end{equation}
where $U$ is a $3\times3$ unitary mixing matrix.
In this scheme there are two independent mass-squared differences,
$ \Delta{m}^2_{21} \equiv m_2^2 - m_1^2 $
and
$ \Delta{m}^2_{31} \equiv m_3^2 - m_1^2 $,
that generate neutrino oscillations
for two different scales of the ratio
$E/L$,
where $E$ is the neutrino energy and $L$ is the distance of propagation.
The solar and atmospheric neutrino anomalies
can be explained in terms of neutrino oscillations
only if
$\Delta{m}^2_{21}$ and $\Delta{m}^2_{31}$
are relevant, respectively,
for the oscillations of solar and atmospheric neutrinos
\cite{SS92,Fogli-Lisi,Goswami,Narayan,BGK96,GKM97}.

In Sections \ref{CHOOZ and solar neutrinos}
and \ref{Atmospheric neutrinos}
we will show that
in the scheme under consideration
the recent results of the CHOOZ experiment \cite{CHOOZ}
imply
that $|U_{e3}|^2\ll1$,
that the oscillations of solar neutrinos are described
by the two-generation formalism
and that
the oscillations of solar and atmospheric neutrinos
depend on different and independent elements of the neutrino
mixing matrix,
\emph{i.e.}
they decouple.
In Section \ref{Atmospheric neutrinos}
it is shown that if not only
$|U_{e3}|^2\ll1$
but also
$|U_{e3}|\ll1$,
then
the oscillations of atmospheric neutrinos do not depend
on matter effects and are described by the two-generation formalism.
In this case,
with an appropriate identification of the mixing parameters,
the formalism
used in the two-generation analyses of solar and atmospheric neutrino data
is appropriate also in the case of three-neutrino mixing
and the results of these analyses
provide direct information on the mixing parameters
of three neutrinos.
In Section \ref{The mixing matrix}
we discuss the implications of the present
solar and atmospheric neutrino data
for the values of the elements of the neutrino mixing matrix.
As shown in Section \ref{Accelerator long-baseline experiments},
the hypothesis $|U_{e3}|\ll1$
could be tested
by future accelerator long-baseline neutrino oscillation
experiments.

In this paper we do not consider
the indication in favor of neutrino oscillations
obtained by the LSND experiment \cite{LSND},
which needs the enlargement of the neutrino mixing scheme
with the introduction of a sterile neutrino
(see \cite{BGG96,OY96}).
A possible decoupling of solar and atmospheric neutrino oscillations
in this enlarged scheme
will be discussed elsewhere.

\section{CHOOZ and solar neutrinos}
\label{CHOOZ and solar neutrinos}

The first reactor long-baseline
neutrino oscillation experiment CHOOZ
found no evidence for
neutrino oscillations \cite{CHOOZ}.
The CHOOZ collaboration has published an exclusion curve
in the plane of the two-generation mixing parameters
which shows that
\begin{equation}
\sin^22\vartheta_{\mathrm{CHOOZ}} \leq 0.18
\quad \mbox{for} \quad
\Delta{m}^2_{\mathrm{CHOOZ}} \gtrsim 10^{-3} \, \mathrm{eV}^2
\,.
\label{51}
\end{equation}

In the three-neutrino scheme under consideration
these parameters are given by
(see \cite{BGK96,GKM97})
\begin{equation}
\sin^22\vartheta_{\mathrm{CHOOZ}} = 4 |U_{e3}|^2 ( 1 - |U_{e3}|^2 )
\quad \mbox{and} \quad
\Delta{m}^2_{\mathrm{CHOOZ}} = \Delta{m}^2_{31}
\,.
\label{52}
\end{equation}
If
$ \Delta{m}^2_{31} > 10^{-3} \, \mathrm{eV}^2 $,
as indicated by the solution of the atmospheric neutrino anomaly
\cite{Kam-atm},
the upper bound (\ref{52}) on $\sin^22\vartheta_{\mathrm{CHOOZ}}$
implies that
\begin{equation}
|U_{e3}|^2 \leq 5 \times 10^{-2}
\quad \mbox{or} \quad
|U_{e3}|^2 \geq 0.95
\,.
\label{02}
\end{equation}

A large value of $|U_{e3}|^2$
does not allow to explain the solar neutrino problem
with neutrino oscillations (see \cite{BGKM97}).
Indeed,
in the scheme under consideration
the averaged survival probability of solar electron neutrinos
is given by \cite{SS92}
\begin{equation}
P_{\nu_e\to\nu_e}^{\mathrm{sun}}(E)
=
\left(
1
-
|U_{e3}|^2
\right)^2
P_{\nu_e\to\nu_e}^{(1,2)}(E)
+
|U_{e3}|^4
\,,
\label{35}
\end{equation}
where
$E$ is the neutrino energy
and
$ P_{\nu_e\to\nu_e}^{(1,2)}(E) $
is the two-generation survival probability of solar $\nu_{e}$'s
which depends on
\begin{equation}
\Delta{m}^2_{\mathrm{sun}}
=
\Delta{m}^2_{21}
\quad \mbox{and} \quad
\sin\vartheta_{\mathrm{sun}}
=
\frac{ |U_{e2}| }{ \sqrt{ 1 - |U_{e3}|^2 } }
\,.
\label{36}
\end{equation}
The expression (\ref{35}) implies that
$ P_{\nu_e\to\nu_e}^{\mathrm{sun}}(E) \geq |U_{e3}|^4 $
and
for
$ |U_{e3}|^2 \geq 0.95 $
we have
$ P_{\nu_e\to\nu_e}^{\mathrm{sun}}(E) \geq 0.90 $.
With such a high and practically constant value of
$P_{\nu_e\to\nu_e}^{\mathrm{sun}}(E)$
it is not possible to explain the suppression of the solar $\nu_e$ flux
measured by all experiments
(Homestake \cite{Homestake},
Kamiokande \cite{Kam-sun},
GALLEX \cite{GALLEX},
SAGE \cite{SAGE}
and
SuperKamiokande \cite{SK-sun})
with respect to that predicted by the Standard Solar Model
\cite{Bahcall,Saclay,CDF}.

Therefore,
we are led to the conclusion that
the results of the CHOOZ experiment imply that
$|U_{e3}|^2$ is small:
\begin{equation}
|U_{e3}|^2 \leq 5 \times 10^{-2}
\,.
\label{04}
\end{equation}

Furthermore, from Eqs.(\ref{35}) and (\ref{36})
one can see that for such small values of $|U_{e3}|^2$
we have
\begin{equation}
P_{\nu_e\to\nu_e}^{\mathrm{sun}}(E)
\simeq
P_{\nu_e\to\nu_e}^{(1,2)}(E)
\quad \mbox{and} \quad
\sin\vartheta_{\mathrm{sun}} \simeq |U_{e2}|
\,.
\label{05}
\end{equation}
Hence the two-generation analyses of the solar neutrino data
are appropriate
in the three-neutrino scheme with a mass hierarchy
and they give information on the values of
\begin{equation}
\Delta{m}^2_{21} = \Delta{m}^2_{\mathrm{sun}}
\quad \mbox{and} \quad
|U_{e2}| \simeq \sin\vartheta_{\mathrm{sun}}
\,.
\label{71}
\end{equation}

According to the most recent analysis of the solar neutrino data
\cite{FLM97},
which include preliminary data from SuperKamiokande
\cite{SK-sun},
the ranges of the mixing parameters allowed at 90\% CL
for the small and large mixing angle MSW \cite{MSW} solutions
and for the vacuum oscillation solution are,
respectively,
\begin{eqnarray}
4 \times 10^{-6} \, \mathrm{eV}^2
\lesssim
\Delta{m}^2_{\mathrm{sun}}
\lesssim
1.2 \times 10^{-5} \, \mathrm{eV}^2
\,,
\null & \null \quad \null & \null
3 \times 10^{-3}
\lesssim
\sin^22\vartheta_{\mathrm{sun}}
\lesssim
1.1 \times 10^{-2}
\,,
\label{23}
\\
8 \times 10^{-6} \, \mathrm{eV}^2
\lesssim
\Delta{m}^2_{\mathrm{sun}}
\lesssim
3.0 \times 10^{-5} \, \mathrm{eV}^2
\,,
\null & \null \quad \null & \null
0.42
\lesssim
\sin^22\vartheta_{\mathrm{sun}}
\lesssim
0.74
\,,
\label{24}
\\
6 \times 10^{-11} \, \mathrm{eV}^2
\lesssim
\Delta{m}^2_{\mathrm{sun}}
\lesssim
1.1 \times 10^{-10} \, \mathrm{eV}^2
\,,
\null & \null \quad \null & \null
0.70
\lesssim
\sin^22\vartheta_{\mathrm{sun}}
\leq
1
\,.
\label{25}
\end{eqnarray}
Therefore,
taking into account that $|U_{e3}|^2$ is small
and assuming that
$|U_{e2}|\leq|U_{e1}|$
(this choice is necessary only for the MSW solutions),
we have
\begin{eqnarray}
|U_{e1}|
\simeq
1
\,,
\quad
|U_{e2}|
\simeq
0.03 - 0.05
\null & \null \quad \null & \null
\mbox{(small mixing MSW)}
\,,
\label{31}
\\
|U_{e1}|
\simeq
0.87 - 0.94
\,,
\quad
|U_{e2}|
\simeq
0.35 - 0.49
\null & \null \quad \null & \null
\mbox{(large mixing MSW)}
\,,
\label{32}
\\
|U_{e1}|
\simeq
0.71 - 0.88
\,,
\quad
|U_{e2}|
\simeq
0.48 - 0.71
\null & \null \quad \null & \null
\mbox{(vacuum oscillations)}
\,.
\label{33}
\end{eqnarray}
Notice that these ranges are statistically rather stable.
For example,
the range of
$\sin^22\vartheta_{\mathrm{sun}}$
allowed at 99\% CL
in the case of the large mixing angle MSW solution is
$
0.36
\lesssim
\sin^22\vartheta_{\mathrm{sun}}
\lesssim
0.85
$
\cite{FLM97},
which imply
$
|U_{e1}|
\simeq
0.83 - 0.95
$,
$
|U_{e2}|
\simeq
0.32 - 0.55
$
(confront with Eq.(\ref{32})).

\section{Atmospheric neutrinos}
\label{Atmospheric neutrinos}

The evolution equation for the flavor amplitudes
$\psi_{\alpha}$ ($\alpha=e,\mu,\tau$)
of atmospheric neutrinos
propagating in the interior of the Earth
can be written as
(see \cite{KP89,GKM97})
\begin{equation}
i
\,
\frac{ \mathrm{d} }{ \mathrm{d}t }
\,
\Psi
=
\frac{ 1 }{ 2 \, E }
\left(
U
\,
M^2
\,
U^{\dagger}
+
A
\right)
\Psi
\,,
\label{13}
\end{equation}
with
\begin{equation}
\Psi
\equiv
\left(
\begin{array}{l}
\psi_{e}
\\
\psi_{\mu}
\\
\psi_{\tau}
\end{array}
\right)
\,,
\quad
M^2
\equiv
\mathrm{diag}( 0 , \Delta{m}^2_{21} , \Delta{m}^2_{31} )
\,,
\quad
A
\equiv
\mathrm{diag}( A_{CC} , 0 , 0 )
\,,
\label{14}
\end{equation}
and
$ A_{CC} \equiv 2 E V_{CC} $,
where
$ V_{CC} = \sqrt{2} G_{F} N_{e} $
is the charged-current effective potential which depends on the
electron number density $N_{e}$ of the medium
($G_{F}$ is the Fermi constant
and
for anti-neutrinos
$ A_{CC} $
must be replaced by
$ \bar{A}_{CC} = - A_{CC} $).
If the squared-mass difference
$\Delta{m}^2_{21}$
is relevant for the explanation of
the solar neutrino problem,
we have
\begin{equation}
\frac{ \Delta{m}^2_{21} \, R_{\oplus} }{ 2 \, E }
\ll
1
\,,
\label{15}
\end{equation}
where
$ R_{\oplus} = 6371 \, \mathrm{Km} $
is the radius of the Earth.
Notice, however, that caution
is needed for low-energy atmospheric neutrinos if
$ \Delta{m}^2_{21} \gtrsim 10^{-5} \, \mathrm{eV}^2 $,
as in the case of the large mixing angle MSW solution
of the solar neutrino problem
and marginally in the case of the small mixing angle MSW solution
(see Eqs.(\ref{24}) and (\ref{23})).
Indeed,
if
$ \Delta{m}^2_{21} \gtrsim 10^{-5} \, \mathrm{eV}^2 $
we have
$ \Delta{m}^2_{21} R_{\oplus} / 2 E \ll 1 $
only for
$ E \gg 150 \, \mathrm{MeV} $.
In this case, in order to get information
on the three-neutrino mixing matrix
with a two-generation analysis
it is necessary to analyze the atmospheric neutrino data
with a cut in energy such that
$ \Delta{m}^2_{21} R_{\oplus} / 2 E \ll 1 $.
In order to be on the safe side,
when we will consider the case of
the MSW solutions of the solar neutrino problem
we will take into account the information
obtained from the two-generation fit of the preliminary
SuperKamiokande multi-GeV data alone
(Figure 11 of \cite{Valencia}).

The inequalities (\ref{15}) imply that
the phase generated by
$\Delta{m}^2_{21}$
can be neglected for atmospheric neutrinos
and $M^2$
can be approximated with
\begin{equation}
M^2
\simeq
\mathrm{diag}(0,0,\Delta{m}^2_{31})
\,.
\label{16}
\end{equation}
In this case
(taking into account that
the phases of the matrix elements $U_{\alpha3}$
can be included in the charged lepton fields)
we have
\begin{equation}
(
U
\,
M^2
\,
U^{\dagger}
)_{\alpha'\alpha}
\simeq
\Delta{m}^2_{31}
\,
|U_{\alpha'3}|
\,
|U_{\alpha3}|
\,.
\label{17}
\end{equation}
Comparing this expression with Eqs.(\ref{05}) and (\ref{71}),
one can see that
the oscillations of solar and atmospheric neutrinos
depend on different and independent
$\Delta{m}^2$'s
and on different and independent 
elements of the mixing matrix,
\emph{i.e.} they are decoupled.
Strictly speaking
$|U_{e2}|$ in Eqs.(\ref{05}) and (\ref{71})
is not independent from $|U_{e3}|$
because of the unitarity constraint
$|U_{e1}|^2+|U_{e2}|^2+|U_{e3}|^2=1$,
but the limit (\ref{04}) on
$|U_{e3}|^2$
implies that its contribution to 
the unitarity constraint is negligible.

Hence, we have shown that the smallness of $|U_{e3}|^2$
inferred from the results of the CHOOZ experiment
imply that the oscillations of solar and atmospheric neutrinos
are decoupled.

From Eqs.(\ref{13}) and (\ref{17}) one can see that
unless $|U_{e3}|\ll1$,
the evolution equations of the atmospheric
electron neutrino amplitude $\psi_e$
and those of the muon and tau neutrino amplitudes
$\psi_\mu$ and $\psi_\tau$
are coupled.
In this case matter effects can contribute
to the dominant $\nu_\mu\to\nu_\tau$
oscillations
(see \cite{FLMM97})
and
the atmospheric neutrino data must be analyzed
with the three generation evolution equation (\ref{13}). 

From the results of the CHOOZ experiment it follows that
the quantity $|U_{e3}|^2$ is small
and satisfy the inequality (\ref{04})
(for $ \Delta{m}^2_{31} \gtrsim 10^{-3} \, \mathrm{eV}^2 $).
However,
the upper bound for $|U_{e3}|$ implied by Eq.(\ref{04})
is not very strong:
$ |U_{e3}| < 0.22 $.
In the following
we will assume that
not only
$|U_{e3}|^2\ll1$,
but also the element
$|U_{e3}|$
that connects the first and third generations is small:
$|U_{e3}|\ll1$
(let us remind that in the quark sector
$ 2 \times 10^{-2} \leq |V_{ub}| \leq 5 \times 10^{-2} $).
We will consider the other elements of the mixing matrix
as free parameters
and we will see that these parameters can be determined
by the two-neutrino analyses of the solar and atmospheric
neutrino data.
In Section \ref{Accelerator long-baseline experiments}
it will be shown that
the hypothesis
$|U_{e3}|\ll1$
can be tested in future long-baseline neutrino oscillation experiments.

If
$|U_{e3}|\ll1$,
for the evolution operator in Eq.(\ref{13})
we have the approximate expression
\begin{equation}
U
\,
M^2
\,
U^{\dagger}
+
A
\simeq
\Delta{m}^2_{31}
\left(
\begin{array}{ccc}
\frac{ A_{CC} }{ \Delta{m}^2_{31} }
&
0
&
0
\\
0
&
|U_{\mu3}|^2
&
|U_{\mu3}| |U_{\tau3}|
\\
0
&
|U_{\tau3}| |U_{\mu3}|
&
|U_{\tau3}|^2
\end{array}
\right)
\label{19}
\end{equation}
which shows that the evolution of
$\nu_e$
is decoupled from the evolution of
$\nu_\mu$ and $\nu_\tau$.
Thus, the survival probability of atmospheric $\nu_e$'s
is equal to one
and
$\nu_\mu\to\nu_\tau$
transitions are independent from matter effects and
are described by a two-generation formalism.
In this case, the two-generation analyses of
the atmospheric neutrino data in terms of $\nu_\mu\to\nu_\tau$
are appropriate
in the three-neutrino scheme under consideration and
yield information on the values of the parameters
\begin{equation}
\Delta{m}^2_{31} = \Delta{m}^2_{\mathrm{atm}}
\quad \mbox{and} \quad
|U_{\mu3}| = \sin\vartheta_{\mathrm{atm}}
\,.
\label{72}
\end{equation}

According to a recent analysis \cite{Valencia}
of the atmospheric neutrino data,
the ranges of
$\Delta{m}^2_{\mathrm{atm}}$
and
$\sin^22\vartheta_{\mathrm{atm}}$
for $\nu_\mu\to\nu_\tau$ oscillations
allowed at 90\% CL
by the SuperKamiokande multi-GeV data
and by all data are, respectively,
\begin{eqnarray}
4 \times 10^{-4} \, \mathrm{eV}^2
\lesssim
\Delta{m}^2_{\mathrm{atm}}
\lesssim
8 \times 10^{-3} \, \mathrm{eV}^2
\,,
\null & \null \quad \null & \null
0.72
\lesssim
\sin^22\vartheta_{\mathrm{atm}}
\leq
1
\,,
\label{61}
\\
4 \times 10^{-4} \, \mathrm{eV}^2
\lesssim
\Delta{m}^2_{\mathrm{atm}}
\lesssim
6 \times 10^{-3} \, \mathrm{eV}^2
\,,
\null & \null \quad \null & \null
0.76
\lesssim
\sin^22\vartheta_{\mathrm{atm}}
\leq
1
\,.
\label{62}
\end{eqnarray}
Thus, assuming that
$ |U_{\mu3}| \leq |U_{\tau3}| $
and taking into account the comments after Eq.(\ref{15}),
we have
\begin{eqnarray}
|U_{\mu3}|
\simeq
0.49 - 0.71
\,,
\quad
|U_{\tau3}|
\simeq
0.71 - 0.87
\null & \null \quad \null & \null
\mbox{(MSW)}
\,,
\label{63}
\\
|U_{\mu3}|
\simeq
0.51 - 0.71
\,,
\quad
|U_{\tau3}|
\simeq
0.71 - 0.86
\null & \null \quad \null & \null
\mbox{(vacuum oscillations)}
\,.
\label{64}
\end{eqnarray}
As in the case of the ranges (\ref{31})--(\ref{33}),
also the ranges (\ref{63})--(\ref{64})
are statistically rather stable.
For example,
the range of
$\sin^22\vartheta_{\mathrm{sun}}$
allowed at 99\% CL
by all the atmospheric neutrino data is
$
0.66
\lesssim
\sin^22\vartheta_{\mathrm{sun}}
\leq
1
$
\cite{Valencia},
which imply
$
|U_{\mu3}|
\simeq
0.46 - 0.71
$,
$
|U_{\tau}|
\simeq
0.71 - 0.89
$
(confront with Eq.(\ref{64})).

\section{The mixing matrix}
\label{The mixing matrix}

Taking into account the unitarity of the mixing matrix,
the information in
Eqs.(\ref{31})--(\ref{33}) and (\ref{63})--(\ref{64}),
together with the assumption
$|U_{e3}|\ll1$,
allow to infer the allowed ranges for the values of
$|U_{\mu1}|$, $|U_{\mu2}|$, $|U_{\tau1}|$ and $|U_{\tau2}|$.
The simplest way to do it is
to start from the Maiani parameterization
of a $3\times3$ mixing matrix \cite{Maiani}:
\begin{equation}
U
=
\left(
\begin{array}{ccc}
c_{12}
c_{13}
&
s_{12}
c_{13}
&
s_{13}
\\
-
s_{12}
c_{23}
-
c_{12}
s_{23}
s_{13}
&
c_{12}
c_{23}
-
s_{12}
s_{23}
s_{13}
&
s_{23}
c_{13}
\\
s_{12}
s_{23}
-
c_{12}
c_{23}
s_{13}
&
-
c_{12}
s_{23}
-
s_{12}
c_{23}
s_{13}
&
c_{23}
c_{13}
\end{array}
\right)
\,,
\label{43}
\end{equation}
where
$ c_{ij} \equiv \cos\vartheta_{ij} $
and
$ s_{ij} \equiv \sin\vartheta_{ij} $
and
we have omitted possible CP-violating phases
(one for Dirac neutrinos and three for Majorana neutrinos)
on which there is no information.

A very small $|U_{e3}|$ implies that $|s_{13}|\ll1$.
In this case we have
\begin{equation}
U
\simeq
\left(
\begin{array}{ccc}
c_{12}
&
s_{12}
&
\ll 1
\\
-
s_{12}
c_{23}
&
c_{12}
c_{23}
&
s_{23}
\\
s_{12}
s_{23}
&
-
c_{12}
s_{23}
&
c_{23}
\end{array}
\right)
\,.
\label{47}
\end{equation}
Using the information on
$|s_{12}|\simeq|U_{e2}|$
and
$|s_{23}|\simeq|U_{\mu3}|$
given by Eqs.(\ref{31})--(\ref{33}) and (\ref{63})--(\ref{64}),
for the moduli of the elements of the mixing matrix
we obtain:
\begin{eqnarray}
\mbox{Small mixing MSW:}
\null & \null \quad \null & \null
\left(
\begin{array}{ccc}
\simeq 1
&
0.03 - 0.05
&
\ll 1
\\
0.02 - 0.05
&
0.71 - 0.87
&
0.49 - 0.71
\\
0.01 - 0.04
&
0.48 - 0.71
&
0.71 - 0.87
\end{array}
\right)
\,,
\label{44}
\\
\mbox{Large mixing MSW:}
\null & \null \quad \null & \null
\left(
\begin{array}{ccc}
0.87 - 0.94
&
0.35 - 0.49
&
\ll 1
\\
0.25 - 0.43
&
0.61 - 0.82
&
0.49 - 0.71
\\
0.17 - 0.35
&
0.42 - 0.66
&
0.71 - 0.87
\end{array}
\right)
\,,
\label{45}
\\
\mbox{Vacuum oscillations:}
\null & \null \quad \null & \null
\left(
\begin{array}{ccc}
0.71 - 0.88
&
0.48 - 0.71
&
\ll 1
\\
0.34 - 0.61
&
0.50 - 0.76
&
0.51 - 0.71
\\
0.24 - 0.50
&
0.36 - 0.62
&
0.71 - 0.86
\end{array}
\right)
\,.
\label{46}
\end{eqnarray}
Let us remark that
in the case of the small mixing angle MSW solution
of the solar neutrino problem
$|U_{e3}|\ll1$
could be of the same order of magnitude as $|U_{e2}|$.

It is interesting to notice that,
because of the large mixing of
$\nu_\mu$ and $\nu_\tau$
with $\nu_2$,
the transitions of solar $\nu_e$'s
in
$\nu_\mu$'s and $\nu_\tau$'s
are of comparable magnitude.
However,
this phenomenon
and the values of the entries
in the
$(\nu_\mu,\nu_\tau)$--$(\nu_1,\nu_2)$
sector of the mixing matrix
cannot be checked with solar neutrino experiments
because the low-energy
$\nu_\mu$'s and $\nu_\tau$'s
coming from the sun can be detected only with neutral-current
interactions,
which are flavor-blind.
Moreover,
it will be very difficult to check
the values of
$|U_{\mu1}|$, $|U_{\mu2}|$, $|U_{\tau1}|$ and $|U_{\tau2}|$
in laboratory experiments
because of the smallness of $m_2$.

In the derivation of Eqs.(\ref{44})--(\ref{46})
we have assumed that
$|U_{e2}|\leq|U_{e1}|$
and
$|U_{\mu3}|\leq|U_{\tau3}|$.
The other possibilities,
$|U_{e2}|\geq|U_{e1}|$
and
$|U_{\mu3}|\geq|U_{\tau3}|$,
are equivalent, respectively,
to an exchange of the first and second columns
and
to an exchange of the second and third rows
in the matrices (\ref{44})--(\ref{46}).
Unfortunately,
these alternatives are hard to distinguish experimentally
because of the above mentioned difficulty
to measure directly the values of
$|U_{\mu1}|$, $|U_{\mu2}|$, $|U_{\tau1}|$ and $|U_{\tau2}|$.
Only the choice $|U_{e2}|\leq|U_{e1}|$,
which is necessary for the MSW solutions
of the solar neutrino problem,
could be confirmed by the results of the new generation
of solar neutrino experiments
(SuperKamiokande,
SNO,
ICARUS,
Borexino,
GNO
and others
\cite{future-sun})
if they will allow to exclude the vacuum oscillation solution.

\section{Accelerator long-baseline experiments}
\label{Accelerator long-baseline experiments}

Future results
from reactor long-baseline neutrino oscillation experiments
(CHOOZ \cite{CHOOZ},
Palo Verde \cite{PaloVerde},
Kam-Land \cite{Kam-Land})
could allow to improve the upper bound (\ref{04})
on $|U_{e3}|^2$.
In this section we discuss how
an improvement of this upper bound
could be obtained by future
accelerator long-baseline neutrino oscillation experiments
that are sensitive to $\nu_\mu\to\nu_e$ transitions
(K2K \cite{K2K},
MINOS \cite{MINOS},
ICARUS \cite{ICARUS}
and others \cite{otherLBL}).

If matter effects are not important,
in the scheme under consideration
the parameter
$\sin^22\vartheta_{\mu{e}}$
measured in
$\nu_\mu\to\nu_e$
long-baseline experiments is given by
(see \cite{BGK96,GKM97})
\begin{equation}
\sin^22\vartheta_{\mu{e}}
=
4 |U_{e3}|^2 |U_{\mu3}|^2
\,.
\label{56}
\end{equation}
If accelerator long-baseline neutrino oscillation experiments
will not observe
$\nu_\mu\to\nu_e$ transitions
and will place an upper bound
$
\sin^22\vartheta_{\mu{e}}
\leq
\sin^22\vartheta_{\mu{e}}^{\mathrm{(max)}}
$,
it will be possible to obtain the limit
\begin{equation}
|U_{e3}|^2
\leq
\frac
{ \sin^22\vartheta_{\mu{e}}^{\mathrm{(max)}} }
{ 4 |U_{\mu3}|^2_{\mathrm{(min)}} }
\,,
\label{57}
\end{equation}
where
$|U_{\mu3}|^2_{\mathrm{(min)}}$
is the minimum value of
$|U_{\mu3}|^2$
allowed by the solution of the atmospheric neutrino anomaly
and
by the observation of
$\nu_\mu\to\nu_\tau$
long-baseline transitions.
For example,
taking 
$|U_{\mu3}|^2_{\mathrm{(min)}}=0.25$
(see Eq.(\ref{64}))
we have
$
|U_{e3}|^2
\leq
\sin^22\vartheta_{\mu{e}}^{\mathrm{(max)}}
$.
If a value of
$ \sin^22\vartheta_{\mu{e}}^{\mathrm{(max)}} \simeq 10^{-3} $,
that corresponds to the sensitivity of the ICARUS experiment
for one year of running \cite{ICARUS},
will be reached,
it will be possible to
put the upper bound
$ |U_{e3}| \lesssim 3 \times 10^{-2} $.

The observation of
$\nu_\mu\to\nu_\tau$
transitions in long-baseline experiments
will allow to establish a lower bound for
$|U_{\mu3}|^2$
because
the parameter
$\sin^22\vartheta_{\mu\tau}$
is given in the scheme under consideration by
(see \cite{BGK96,GKM97})
\begin{equation}
\sin^22\vartheta_{\mu\tau}
=
4 |U_{\mu3}|^2 |U_{\tau3}|^2
\,.
\label{81}
\end{equation}
From the unitarity relation
$|U_{e3}|^2+|U_{\mu3}|^2+|U_{\tau3}|^2=1$
it follows that
an experimental lower bound
$\sin^22\vartheta_{\mu\tau}\geq\sin^22\vartheta_{\mu\tau}^{\mathrm{(min)}}$
allows to constraint
$|U_{\mu3}|^2$
in the range
\begin{equation}
\frac{1}{2}
\left( 1 - \sqrt{ 1 - \sin^22\vartheta_{\mu\tau}^{\mathrm{(min)}} } \right)
\leq
|U_{\mu3}|^2
\leq
\frac{1}{2}
\left( 1 + \sqrt{ 1 - \sin^22\vartheta_{\mu\tau}^{\mathrm{(min)}} } \right)
\,.
\label{82}
\end{equation}
If
$\sin^22\vartheta_{\mu\tau}^{\mathrm{(min)}}$
is found to be close to one,
as suggested by the solution
of the atmospheric neutrino problem
(see Eqs.(\ref{61}) and (\ref{62})),
the lower bound
$
|U_{\mu3}|^2_{\mathrm{(min)}}
=
\frac{1}{2}
\left( 1 - \sqrt{ 1 - \sin^22\vartheta_{\mu\tau}^{\mathrm{(min)}} } \right)
$
is close to $1/2$.

If matter effects are important,
the extraction of an upper bound for
$|U_{e3}|^2$
from the data of
$\nu_\mu\to\nu_e$
accelerator long-baseline experiments
is more complicated.
In this case the probability
of
$\nu_\mu\to\nu_e$
oscillations
is given by
(see \cite{GKM97})
\begin{equation}
P_{\nu_\mu\to\nu_e}
=
\frac{ 4 |U_{e3}|^2 |U_{\mu3}|^2 }
{ \left( 1 - \frac{ A_{CC} }{ \Delta{m}^2_{31} } \right)^2
+ 4 |U_{e3}|^2 \, \frac{ A_{CC} }{ \Delta{m}^2_{31} } }
\,
\sin^2\left(
\frac{ \Delta{m}^2_{31} L }{ 4 E }
\,
\sqrt{ \textstyle \left( 1 - \frac{ A_{CC} }{ \Delta{m}^2_{31} } \right)^2
+ 4 |U_{e3}|^2 \, \frac{ A_{CC} }{ \Delta{m}^2_{31} } }
\right)
\,,
\label{83}
\end{equation}
where $E$ is the neutrino energy and $L$ is the distance of propagation.
This probability depends on the neutrino energy
not only through the explicit $E$
in the denominator of the phase,
but also through the energy dependence of
$ A_{CC} \equiv 2 E V_{CC} $.
For long-baseline neutrino beams
propagating in the mantle of the Earth
the charged-current effective potential 
$ V_{CC} = \sqrt{2} G_{F} N_{e} $
is practically constant:
$ N_{e} \simeq 2 \, N_A \, \mathrm{cm}^{-3} $
($N_A$ is the Avogadro number)
and
$ V_{CC} \simeq 1.5 \times 10^{-13} \, \mathrm{eV} $.

If long-baseline experiments will not observe
$\nu_\mu\to\nu_e$
transitions
(or will find that they have an extremely small probability)
for neutrino energies such that
$A_{CC}\lesssim\Delta{m}^2_{31}$,
it will mean that
$|U_{e3}|^2$
is small
and a fit of the experimental data with the formula (\ref{83})
will yield a stringent upper limit for
$|U_{e3}|^2$
(taking into account
the lower limit $|U_{\mu3}|^2\geq|U_{\mu3}|^2_{\mathrm{(min)}}$
obtained from the solution of the atmospheric neutrino anomaly
and
from the observation of
$\nu_\mu\to\nu_\tau$
long-baseline transitions).
On the other hand,
the non-observation of
$\nu_\mu\to\nu_e$
transitions
for neutrino energies such that
$A_{CC}\gg\Delta{m}^2_{31}$
does not provide any information on
$|U_{e3}|^2$
because in this case the transition probability (\ref{83})
is suppressed by the matter effect.
Hence,
in order to check the hypothesis
$|U_{e3}|\ll1$,
as well as to have some possibility to observe
$\nu_\mu\to\nu_e$
transitions if this hypothesis is wrong,
it is necessary that a
substantial part of the energy spectrum of the neutrino beam
lies below
\begin{equation}
\frac{ \Delta{m}^2_{31} }{ 2 V_{CC} }
\simeq
30 \, \mathrm{GeV}
\left( \frac{ \Delta{m}^2_{31} }{ 10^{-2} \, \mathrm{eV}^2 } \right)
\,.
\label{84}
\end{equation}
This requirement will be satisfied in the
accelerator long-baseline experiments under preparation
(K2K \cite{K2K},
MINOS \cite{MINOS},
ICARUS \cite{ICARUS}
and others \cite{otherLBL})
if
$\Delta{m}^2_{31}$
is not much smaller than
$ 10^{-2} \, \mathrm{eV}^2 $.

\section{Conclusions}
\label{Conclusions}

We have considered the scheme of mixing of three neutrinos
with the mass hierarchy (\ref{01})
and with $\Delta{m}^2_{21}$ and $\Delta{m}^2_{31}$
relevant, respectively,
for the oscillations of solar and atmospheric neutrinos.

We have shown that
in the framework of this scheme
the recent results of the CHOOZ experiment \cite{CHOOZ}
imply that $|U_{e3}|^2$ is small,
the oscillations of solar neutrinos are described by the two-generation
formalism
and
the oscillations of solar and atmospheric neutrinos
depend on different and independent elements of the neutrino
mixing matrix,
\emph{i.e.}
they decouple.
We have also shown that if not only
$|U_{e3}|^2\ll1$
but also
$|U_{e3}|\ll1$,
then
the oscillations of atmospheric neutrinos
do not depend on matter effects and
are also described by the two-generation
formalism.
In this case,
with the identifications (\ref{71}) and (\ref{72})
the two-generation analyses of solar and atmospheric neutrino data
provide direct information on the mixing parameters
of three neutrinos
(see Eqs.(\ref{44})--(\ref{46})).

Let us notice that the independence of the oscillations of atmospheric neutrinos
from matter effects can be checked
by comparison of the transition probabilities of neutrinos and antineutrinos
in future accelerator long-baseline experiments.

If future results from reactor
(CHOOZ \cite{CHOOZ},
Palo Verde \cite{PaloVerde},
Kam-Land \cite{Kam-Land})
and accelerator
(K2K \cite{K2K},
MINOS \cite{MINOS},
ICARUS \cite{ICARUS}
and others \cite{otherLBL})
long-baseline neutrino oscillation experiments
will confirm and improve the upper bound
(\ref{04}) for $|U_{e3}|^2$
obtained from the
first results of the CHOOZ experiment \cite{CHOOZ},
the indications in favor of a decoupling
of solar and atmospheric neutrino oscillations
and of their accurate description by the two-generation formalism
will be strengthened.
In this case
the distinction of the three allowed solutions of the solar neutrino problem
(small and large mixing angle MSW and vacuum oscillations),
which is a goal of the new generation of solar neutrino experiments
(SuperKamiokande,
SNO,
ICARUS,
Borexino,
GNO
and others
\cite{future-sun})
could provide an indication of the actual values
of the elements of the neutrino mixing matrix
selecting one of the three possibilities (\ref{44})--(\ref{46}).

\acknowledgments

C.G. would like to thank V. Berezinsky
for a stimulating discussion
on neutrino physics and astrophysics.

\end{document}